# Bacatá: Notebooks for DSLs, Almost for Free

## Mauricio Verano Merino[a,d], Jurgen Vinju[a,b], and Tijs van der Storm[b,c]

a   Eindhoven University of Technology, The Netherlands
b   Centrum Wiskunde & Informatica, The Netherlands
c   University of Groningen, The Netherlands
d   Océ Technologies B.V., The Netherlands

**Abstract**   *Context:* Computational notebooks are a contemporary style of *literate programming*, in which users can communicate and transfer knowledge by interleaving executable code, output, and prose in a single rich document. A Domain-Specific Language (DSL) is a software language tailored for an application domain. Usually, DSL users are domain experts that may not have a software engineering background. Therefore, they might not be familiar with Integrated Development Environments (IDEs). In brief, the development of tools that offer different interfaces for interacting with a DSL is relevant.

*Inquiry:* However, DSL designers' resources are limited. We want to leverage General-purpose Languages (GPLs) tooling in the context of DSLs. Computational notebooks are an example of such tools. Then, our main question is: What is an efficient and effective method of designing and implementing notebook interfaces for DSLs? By addressing this question, we might be able to speed up the development of DSL tools, and ease the interaction between end-users and DSLs.

*Approach:* In this paper, we present Bacatá, a mechanism for generating notebook interfaces for external DSLs in a language-parametric fashion. This mechanism is designed in a way in which language engineers can reuse as many language components as possible (e.g., language processors, type checkers, code generators). In addition, we present a Feature-Oriented Domain Analysis that depicts language dependent and language independent features of computational notebooks.

*Knowledge:* Our results show that notebook interfaces generated by Bacatá can be used with little manual configuration. However, there are a few considerations and caveats that should be addressed by language engineers that rely on language design aspects. The creation of a notebook for a DSL with Bacatá becomes a matter of writing the code that wires existing language components in the Rascal language workbench with the Jupyter platform.

*Grounding:* We evaluate Bacatá by generating functional computational notebook interfaces for three different non-trivial DSLs, namely: a small subset of Halide (a DSL for digital image processing), SweeterJS (an extended version of JavaScript), and QL (a DSL for questionnaires). Additionally, it is relevant to generate notebook implementations rather than implementing them manually. To illustrate this, we measured and compared the number of source lines of code that we reused from existing implementations of those languages.

*Importance:* The adoption of notebooks by novice-programmers and end-users has made them very popular in several domains such as exploratory programming, data science, data journalism, and machine learning. Why are they popular? In (data) science, it is essential to make results reproducible as well as understandable. However, notebooks are only available for GPLs. This paper opens the notebook metaphor for DSLs to improve the end-user experience when interacting with code and to increase DSLs adoption.

## The Art, Science, and Engineering of Programming

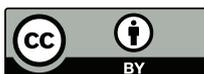

## 1 Introduction

*Computational notebooks* are cell-based documents that allow users to interlace source code and interactive results with prose that explains them. Notebooks have become popular in disciplines such as mathematics, physics, data science, programming education, and machine learning due to the benefits they provide in terms of reproducibility and usability. Currently, there are several dozens of platforms that support the creation of computational notebooks. Wolfram Mathematica was one of the first platforms for notebooks [17][75], yet its appropriation was limited due to their commercial licensing model. Later, in 2014, Project Jupyter [33] developed an open-source notebook platform that has widespread the adoption of the notebook metaphor among different disciplines [29, 47, 55, 56, 60].

Jupyter has millions of users across different disciplines, and there are more than one million public Jupyter notebooks available on GitHub repositories [47]. Jupyter uses language *kernels* to execute source code. A language kernel is a mechanism to support additional languages. By default, Jupyter only supports the *iPython* kernel, but it provides an Application Programming Interface (API) for creating language kernels to support additional languages.

The development of new language kernels is cumbersome. They require the implementation of Jupyter's low-level wire protocol [27], which requires substantial effort. Language Workbenches (LWBs) offer a default set of generic IDE services [13]. Prior research [4, 6, 13, 19, 31, 53, 62] has shown that IDE generation is feasible for Domain-Specific Languages (DSLs). Nevertheless, no support is available for generating language kernels for computational notebooks. The addition of generic language kernels to the LWBs toolbox opens the notebook metaphor for software languages in a generic fashion.

In this paper, we present an extended version of Bacatá [72, 74], and a Feature-Oriented Domain Analysis (FODA) of computational notebooks. Bacatá [1] is a language-parametric kernel generator that hides the complexity of Jupyter's low-level wire protocol. Thus, creating a language kernel for a DSL becomes a matter of writing a few lines of code. Additionally, Bacatá offers a set of generic hooks for registering language-specific services (e.g., syntax highlighting). We included Bacatá as part of the generic IDE services offered by the Rascal LWB [32].

The contributions of this paper are summarized as follows:

- We made a FODA based on 16 computational notebook platforms. Out motivation is to understand their features from the perspective of software language engineering (section 2).
- We present Bacatá-Core, a generic language protocol implemented in Java. This protocol simplifies the development process of Jupyter's language kernels (section 3.2).

---

[1] Available on GitHub at https://github.com/cwi-swat/bacata.

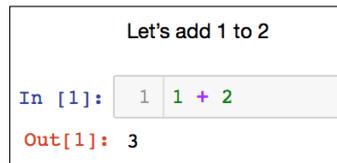

**■ Figure 1** Notebook that calculates the addition of two integers.

- We introduce Bacatá-Rascal, a lightweight bridge between Bacatá-Core and the Rascal LWB. We highlight how to use this bridge to generate language kernels for Rascal DSLs (section 4).
- We evaluate Bacatá by generating language kernels for three different Rascal languages, namely, Halide*(a subset of Halide [51]) a DSL for digital image processing, SweeterJS an extended version of JavaScript, and QL [13] a DSL for defining questionnaires. Moreover, we measure the amount of reused code for generating a language kernel (section 6).

We conclude the paper with a discussion of related work and future research directions (sections 8 and 9).

## 2 Computational Notebooks

Computational notebooks are a contemporary style of literate programming [34] in which users can interleave documentation, source code, and output in a single linear document. Therefore, notebooks enable end-users to teach, learn, and share knowledge.

### 2.1 Anatomy of a Notebook

A computational notebook is a cell-based document, like a spreadsheet with a single column, where cells alternate between computational input, output, and documentation. In section 7 we compare notebooks and spreadsheets. Notebooks have three different types of cells, namely, documentation, input, and output cells. The first is used to write prose. The second contains the executable source code. The last one holds computation results, such as numbers or (interactive) plots and diagrams.

Figure 1 shows an example of a notebook with three cells. The first row is a documentation cell that contains prose text explaining what is going to happen. The second cell (In [1]) displays an input cell where the user has entered the expression 1 + 2 in some programming language. Finally, the last cell (Out [1]) shows the output of evaluating the expression.

Furthermore, notebooks are interactive: readers can tweak input parameters, change code snippets, and observe different ways of representing the output. For instance, changing the expression in In [1] triggers the recomputation of the output cell (Out [1]). More advanced styles of notebooks feature interactive visualizations of results as well, which support interactive exploration of (large) data sets.

Computational notebooks are often persisted as a single document (e.g., Jupyter notebooks are stored in a JSON-format file), which makes sharing easy. Notebook results can be relatively easy reproduced, since all the documentation, source code, and results are part of the same document.

## 2.2 DSLs

DSLs are typically small languages targeted to a specific application domain [41]. Fowler [14] classified them in two styles, *internal DSLs* and *external DSLs*. *Internal DSLs* are also known as embedded DSLs. These DSLs are written inside an existing host language. Due to this, they do not require a separate parser. This also means that the DSL is constrained by the syntax of the host language. In contrast, *external DSLs* do not impose any syntax constraints, yet language engineers must implement a parser for the language. The work we present in this paper is focused on textual external DSLs.

## 2.3 Notebooks for DSLs

Most of the existing language kernels for computational notebooks (e.g., Python, R, Julia), are based on full-fledged programming languages. DSLs, however, are languages tailored to a particular domain. They are designed as a means of communication between domain experts and software engineers. This fact motivates us to explore if it might be useful to develop notebooks for DSLs. Below, we analyze why DSL users and DSL engineers might benefit from interactive notebooks.

**End-user programming.** Unlike General-purpose Programming Languages (GPLs), DSLs are used by domain experts who are not necessarily proficient in software engineering or computer science. Notebooks provide a more friendly interface for interacting with source code and documentation than full-fledged IDEs or basic text editors. Additionally, the fact that notebooks run on ordinary web browsers avoids installation hassle. In summary, notebooks make interaction with code less intimidating.

**Experimentation and simulation.** Computational notebooks deviate from the traditional software development setting where the goal is to build production-quality software, towards a setting where exploration and experimentation take center stage. In the context of DSLs, notebooks allows both language engineers and domain experts to experiment with the language design, enjoy immediate feedback and reproduce previous experiments. However, as notebooks are used for experimentation, they often contain poor quality code [18]. DSL programs written in a notebook can provide input to production-level code generators that create the actual software. Thus, notebooks reinforce the division of labor between domain engineers and application engineers promoted by Domain-Specific Software Engineering (DSE) [5].

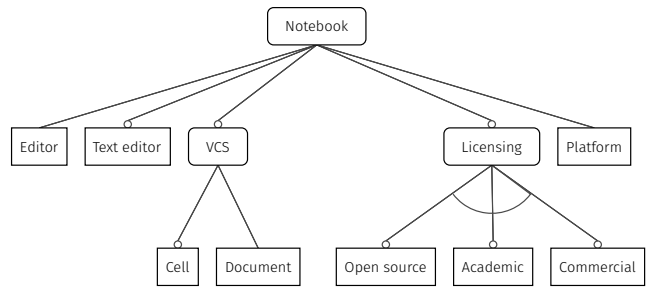

**■ Figure 2** Feature model of computational notebooks.

**DSL education.** DSLs are developed by smaller teams than GPLs like Java or C#. Thus, their development incurs documentation and training costs. Computational notebooks can function as tutorials, providing interactive walk-throughs for a DSL. Thus, notebooks may complement standard forms of documentation (e.g., user guides, reference manuals, API documentation), to allow domain experts to familiarize themselves with a new DSL.

**Language engineering benefits.** There are different engineering trade-offs in the development of DSLs compared to the development of GPLs. DSLs are often developed in-house, by smaller teams, and require a faster design iteration cycle. Notebooks can provide a valuable tool in the language engineer's toolbox for testing and debugging a language implementation. Especially since various language engineering aspects can be exposed as part of the notebook. For instance, notebooks can display the results of different language components (section 6), such as code generators, interpreters, compilers, and type checkers.

### 2.4 Feature-oriented Domain Analysis

Computational notebooks are a form of literate programming [34]. There are different platforms offering several features. Some of these features are language-dependent, and some others are language-independent. We want to offer notebook support for DSLs in a generic fashion. However, first, we need to understand the state-of-the-art of computational notebooks, and then find if it is feasible to automatically generate some notebook features based on existing DSL components. Thus, we conducted a Feature-Oriented Domain Analysis (FODA) [28]. We studied 16 computational notebook platforms (the list of tools and the mapping is shown in appendix A and tables 2 and 3, respectively). The result of the FODA is a feature model (figures 2 to 4).

In the feature model, we use two kinds of features, *mandatory* and *optional*. The first type is used for common features (depicted as a box in figures 2 to 4), and the second type for unique features (depicted as a box with a blank circle on top in figures 2 to 4). The root node in figure 2 represents a notebook. All the notebook's children nodes are described below.

**Editor.** The notebook editor is the Graphical User Interface (GUI) for creating computational narratives. Figure 3 shows a more detailed view of the editor. The Editor

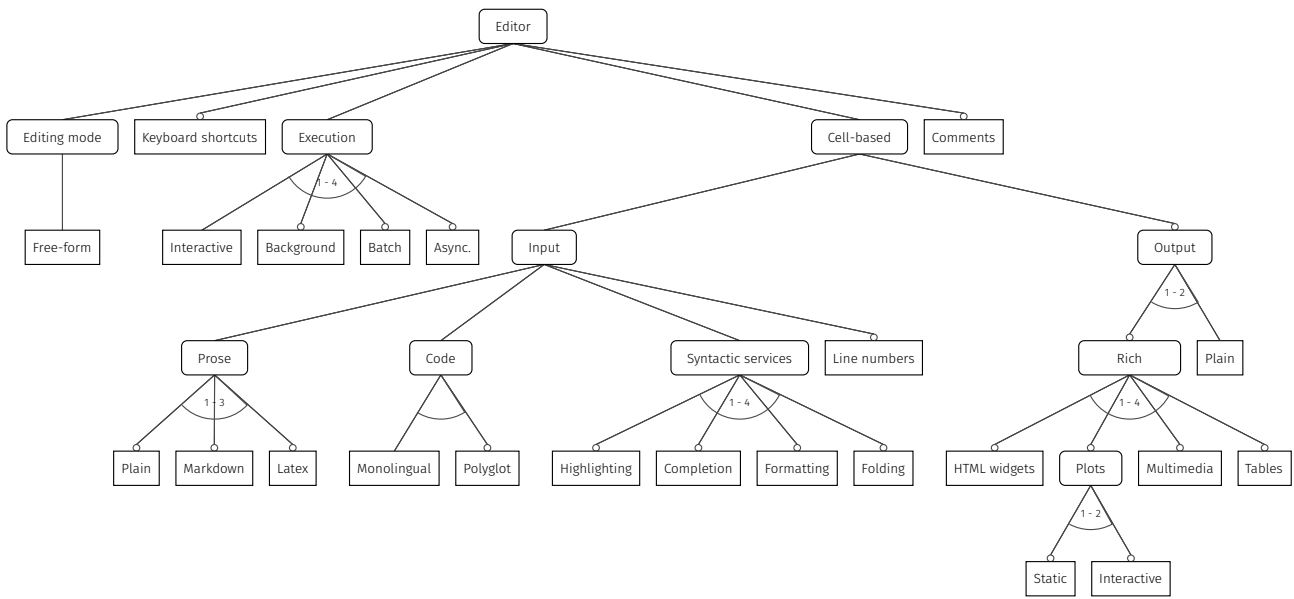

■ **Figure 3** Detailed view of the editor feature.

feature has five sub-features, namely *Editing mode*, *Keyboard shortcuts*, *Execution*, *Cell-based*, and *Comments*.

**Editing mode.** The only editing mode supported is *Free-form* in which the users freely edit both code and documentation cells.

**Keyboard shortcuts.** This feature increases the productivity of experienced users. Not all platforms consider this a must-have feature (table 3).

**Execution.** This feature refers to the process of evaluating input code and rendering a response. We found four different modes. The default mode is *Interactive*, which means that the response is immediate (not necessarily live [54, 70]) after pressing the execution button. The other execution modes, included as optional features, are *Background*, *Batch*, and *Asynchronous*. Support for these execution modes is domain-specific. This means that there are domains where these types of processing are essential to satisfy functional and non-functional requirements. For instance, in data science, the analysis of a big corpus of data is scheduled in batch, due to the amount of time and resources required for its processing.

**Cell-based.** Notebooks have *Input* and *Output* cells. The first can be either *Prose* in several formats (e.g., plain, markdown, LaTeX) or *Code*. There are two types of *Code* cells, *monolingual* and *polyglot*. A monolingual notebook supports one language, while a *polyglot* notebook [43, 45, 49, 61, 67] allows users to execute *Code* cells using different languages within the same notebook. Traditional IDEs offer a set of *Syntactic editor services* [13] to improve the user's productivity and programming experience. The syntactic services offered by notebook are *Syntax highlighting*, *Tab-completion*, *Formatting*, and *Folding*. Finally, *Line numbers* are helpful for error handling and code review. Conversely, output cells are used to display language results (e.g., the output of executing some *Code* cell). These results can be displayed either in *Rich* or *Plain* format. Notebooks support the following rich media elements: *HTML widgets*, *Plots* with either static or dynamic content, *Multimedia* (e.g., images, animations), and *Tables*.

**Comments.** Some platforms allow users to add comments to the notebook itself, and not only as code comments.

**Text editor.** Text editors are included as part of a notebook platform to edit any file. This feature optional because it is not present in all the platforms.

**Version control system (VCS).** Version control is fundamental for managing changes, yet there is not a standardized way of tracking changes in a notebook environment. We found two ways of doing versioning of notebooks, either *Document-oriented* or *Cell-oriented*. The former keeps track of all the changes at a document level, there's no notion of cells. The latter keeps track of all the changes at a cell level, which means the VCS show modifications per cell and not per document.

**Licensing.** There are three licensing models used by the studied platforms, *Open-source*, *Academic*, and *Commercial*.

**Platform.** It embodies all the additional components of a notebook platform, beyond the editor. We divided the platform into five sub-features, as shown in figure 4.

We explain each Platform sub-feature below.

**Deployment.** There are two deployment models in notebook platforms, *Standalone* and *Software as a Service* (SaaS). In the first model, users require an infrastructure to run the platform, so they are responsible for all maintenance activities (e.g., updates, security). In the second model, users do not require an infrastructure. They can use the notebooks straight out of the box, yet the only requirement is a computer with an internet connection and a web browser.

**Extensibility.** Notebook platforms allow developers to enhance their default behavior through a set of APIs. There are two ways of extending these platforms, either by integrating *Third-party* applications or services, or by building *Extensions*.

**Programming language.** Notebook platforms can support either one or multiple programming languages. In this context, we consider that a platform supports multiple programming languages if it allows users to create notebooks with different programming languages. For instance, MATLAB live editor [22] only supports MATLAB as a programming language, while Jupyter supports several programming languages [24].

**Shareability.** Notebooks can be easily shared to have multiple users working on the same notebook. Each user may be focused on different cells in the document. The sharing can be done *online* or *offline*. On the one hand, online sharing means that two or more users can modify the same notebook at the same time, while visualizing modifications made by other users. On the other hand, offline sharing does not allow the modification of the notebook simultaneously. Instead, a user works on his own version of the document. This feature encourages collaboration among different people. Sharing capabilities are supported by the single document metaphor adopted by notebooks.

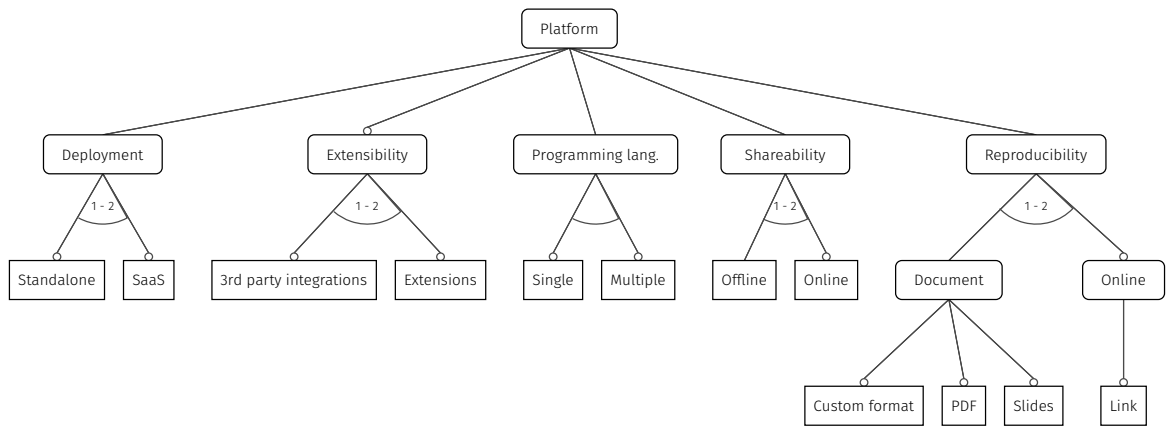

■ **Figure 4**  Detailed view of the platform feature.

**Reproducibility.** Notebooks are often used in scientific contexts, so reproducibility is essential for peer review, validation, and verification. Notebooks contain both the explanation and development of a scientific result and provide the ability to reliably reproduce previous computations using the same data to obtain identical results. There are two different means of sharing a notebook, either by sending the *Document* or by sharing a *Link* that points out to the notebook. In general, there is no standard format for notebooks, so each platform has its own. However, most platforms allow users to export the notebook's content, including or excluding (when privacy is an issue) its results (output cells) in different file formats such as PDF or as slides.

**Summary.**  Looking at the feature model, we can observe that some features are language-dependent, and others are language-independent. The following features are in the first category: highlighting, completion, formatting, folding, and rich media. The other features are orthogonal to the language-specific features and are handled generically by notebook frameworks such as Jupyter.

Apart from rich media output, perhaps, the language-specific features are already part of the standard toolset of LWBs [13], which opens the possibility to reuse language components. In the following section, we present Bacatá, and we describe how we generate Jupyter language kernels using a LWB. Moreover, we demonstrate how language engineers can reuse existing language-specific components within a computational notebook.

## 3 Bacatá

Bacatá is a language-parametric interface between Jupyter and LWBs. It provides a mechanism to generate Jupyter kernels by reusing existing language components such as grammars, parsers, type checkers, code generators, and interpreters. Bacatá also

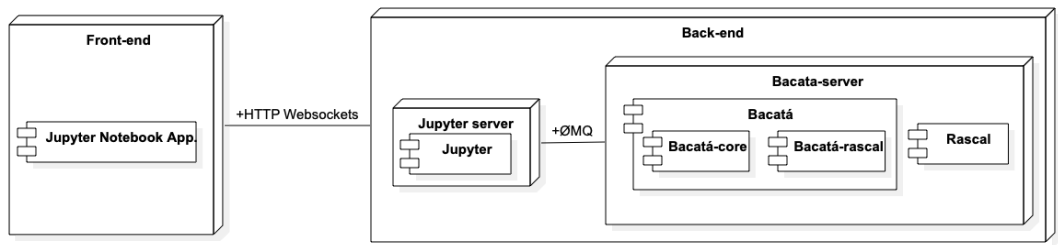

■ **Figure 5** Bacatá's architecture – deployment view.

supports the generation of language-specific components (e.g., syntax highlighters and completors) described in the FODA.

Bacatá's design is based on the notebook features described in the FODA (section 2.4). In the remainder of this section, we detail Bacatá's design and implementation. First, we present its architecture; second, we explain in detail both Bacatá-Core(Bacatá's language service interface (Bacatá-Core) and Bacatá's LWB bridge (Bacatá-Rascal); and third, we show an example of how to generate a language kernel for a toy language using Bacatá.

### 3.1 Architecture

figure 5 shows Bacatá's architecture. The diagram contains two nodes, *Front-end* and *Back-end*. The *Front-end* node represents the user's browser, and it hosts the Jupyter notebook app [23]. The *Back-end* node has two sub-nodes, *Jupyter server* and *Bacatá-server*. The first, hosts Jupyter's logic and services. The latter contains Bacatá's components and the desired LWB (e.g., Rascal). More details about the Bacatá-server node are presented below.

Bacatá has two components, Bacatá-Core and Bacatá-Rascal. Bacatá-Core enables the communication between Jupyter and a LWB. It exposes the ILanguageProtocol, which is a generic language protocol interface (comparable to Microsoft's Language Server Protocol [37]). This interface is independent of Rascal so it can be implemented to work with any LWB. Bacatá-Core is also responsible for collecting the user's code and sending it to the respective language interpreter.

On the contrary, Bacatá-Rascal is Rascal-dependent. It implements the ILanguageProtocol and provides the means for connecting Rascal-based languages to Bacatá-Core. A Kernel type is Bacatá-Rascal's entry point. It is an Algebraic Data Type (ADT) designed to capture language-specific information. This ADT is actively used to complete two essential tasks in the kernel generation. First, it generates Jupyter's language kernel; and second, it is the input for generating language-specific artifacts, such as syntax highlighter modes and tab-completion functions.

Bacatá's language kernels are registered automatically as part of Jupyter's supported languages (in the current environment). Therefore, these language kernels also become available to be used through any of the Jupyter's interfaces [23].

End-users do not interact directly with Bacatá; from their perspective, Bacatá-Core and Bacatá-Rascal are hidden components. They must choose a language kernel from Jupyter's notebook interface, and afterwards they can start interacting with the language. When a language kernel is chosen, Jupyter makes a callback to Bacatá to instantiate a language Read–Eval–Print Loop (REPL). A REPL is a language artifact capable of reading expressions, evaluating them, and printing the result of their evaluation.

### 3.2 Bacatá-Core

Jupyter's mechanism to support new languages is through language kernels. A language kernel is *"a program that runs and introspects the user's code"* [25], and it implements the *wire protocol* [27]. This protocol can be implemented in two ways, by creating a Python wrapper kernel or by hand. The former reuses the IPython [46] kernel infrastructure and is meant to be used by languages with Python bindings (e.g., Hy [21]) [26]. The latter must be implemented by hand in the target language, and gives language engineers full control. In this paper, we focus on the second approach since it gives complete control.

The *wire protocol* is Jupyter's communication protocol, and it is implemented using ZeroMQ sockets [1]. Bacatá-Core is responsible for implementing it at the session and presentation layer in the Open Systems Interconnection (OSI) model [79]. This protocol involves five sockets and more than a dozen messages (appendix B). Each socket and message have a single responsibility, but the right composition allows the interaction between Jupyter's notebook interface and language kernels. Given the number of sockets and messages, its implementation is error-prone and difficult to debug.

To prevent language engineers from having to deal with Jupyter's protocol, Bacatá-Core offers an abstract class called MetaJupyterServer and an interface called ILanguageProtocol (appendix C). MetaJupyterServer hides all the complexity of the wire protocol, including its socket management; while the ILanguageProtocol focuses on the language engineering aspects, such as user's code evaluation and auto-completion, and statements completion.

Bacatá-Core is LWB-agnostic. To use Bacatá with a new LWB, language engineers have to fulfil two conditions. First, they need to implement ILanguageProtocol. Second, they must extend the abstract MetaJupyterServer class (appendix D). Figure 10 in appendix F depicts a simplified class diagram to illustrate how Bacatá can be used by other LWBs. For example, if we would like to use Bacatá with a different LWB (e.g., Spoofax). Assuming the LWB already offers a generic REPL, which is accessible to languages developed within the LWB. First, a language engineer must add Bacatá-Core as a dependency. Second, they must implement the ILanguageProtocol, and the abstract methods of the MetaJupyterServer (appendix D).

In the next section, we discuss implementation details and design decisions of the ILanguageProtocol and MetaJupyterServer. We also show the integration with the Rascal LWB and how to generate language kernels in a language-parametric way.

■ **Listing 1**  REPL type definition.

```
1  data REPL = repl(Result(str)  handler, Completion(str) completor);
2
3  alias  Completion = tuple[int position,  list [str]  suggestions];
4
5  data Result  = text(str  result,  list [Message] messages);
```

## 4    Bacatá-Rascal

Bacatá-Rascal is a language kernel generator for Rascal DSLs. Figure 10 in appendix F depicts a class diagram that illustrates the interaction between Bacatá-Core and Bacatá-Rascal. To enable the communication between a Rascal DSL and Jupyter, we implement a Java class `DSLNotebook` that extends the abstract class `MetaJupyterServer`. `DSLNotebook` implements language-specific methods declared in the `MetaJupyterServer` class, such as code completion and code execution. These methods were selected based on the FODA (section 2.4). The `DSLNotebook` class requires an *interpreter*, which is a generic implementation of the `ILanguageProtocol`. It is used to type, introspect, and evaluate user code for all Rascal DSLs. The interpreter is designed as a language-parametric mechanism for hooking a DSL's REPL into Bacatá. Concretely, the generic interpreter takes as input the DSL's source code path and the REPL's qualified name. Bacatá-Rascal uses this information to load the language and execute user code.

To create a new language kernel with Bacatá, language engineers must define a `REPL` (listing 1) for their language. A `REPL` object wraps the language's interactive interpreter. It is used to evaluate user code (`handler` function) and to complete code fragments (`completor` function). The result type of each function (`completor` and `handler`) is also shown in listing 1, lines 3–5.

The walk-through to create a language kernel with Bacatá is explained below.

1.  Language engineers have to specify language's information through a `Kernel` ADT (listing 2). The `Kernel` defines configuration parameters for obtaining language-specific information (e.g., name and logo) required by Bacatá. Additionally, it has information about relevant resources and configuration parameters for the language implementation.

2.  The language engineer calls `bacata` (listing 4), which has two overloaded function definitions. On the one hand, the first definition (listing 4, line 1) takes an argument value of type `Kernel` (listing 2) and two optional boolean parameters, namely, debug and Docker [40]. The first boolean parameter is used for debugging the language kernel. The second boolean parameter is used to generate a *Dockerfile* that assembles all the required dependencies to run the generated computational notebook (including the generated language kernel). The second `bacata` definition (listing 4, line 2) has an extra parameter, namely `grammar`. We included this parameter to offer syntax highlighting support through CodeMirror modes [8]. We explain syntax highlighting details in section 5.1.

3.  The `bacata` function produces several side-effects. First, Bacatá verifies Jupyter's correct installation and the definition of the required environment variables. Second,

■ **Listing 2**   Kernel type definition.

```
1  data Kernel = kernel(str  languageName, loc projectPath, str replFunction, loc logoPath = |tmp:///|);
```

■ **Listing 3**   Notebook type definition.

```
1  data Notebook = notebook(void () serve, void() stop)
2                 | notebook(void() serve, void() stop, void() logs);
```

■ **Listing 4**   Bacata function.

```
1  Notebook bacata(Kernel kernel, bool debug = false, bool docker=false) {...}
2  Notebook bacata(Kernel kernel, type[&T <: Tree] grammar, bool debug = false, bool docker = false) {...}
3  Notebook bacata(Kernel kernel, Mode mode, bool debug = false, bool docker = false) {...}
```

it generates a JSON serialized dictionary (a.k.a. *kernel.json*) that contains language-specific information (including Bacatá's wiring) and Jupyter's connection details (e.g., ZMQ socket ports). Third, it constructs a value of type `Notebook` (listing 3) that encapsulates either two or three functions (depending on the selected overloaded constructor). The `serve` function starts Jupyter's server, while the `stop` function is used to shut down the server. Moreover, to capture Jupyter's server logs, we use the `logs` function. Fourth, it installs the generated language kernel in the current Jupyter environment. Finally, to obtain an updated version of the front-end, Bacatá automatically recompiles all Jupyter's assets (including DSL-specific artifacts such as generated CodeMirror modes).

## 5   Concrete Example: the CALC Language

So far, we have introduced Bacatá's components; now, we are going to explain how to generate a language kernel using Bacatá for a simple calculator language (CALC). This language already existed, and it was already implemented using the Rascal LWB. However, all the Rascal code could also be written in Java or using another LWB. We present Calc's grammar in listing 5. It consists of commands (listing 5, lines 5-7) and expressions (listing 5, lines 1-5). There are two types of supported commands, namely assignments and expression evaluation. Calc's expressions are variables, numbers, multiplication, and addition. To execute commands, there is an `exec` function (listing 5, line 9) that returns a tuple containing an integer, and a possibly updated environment (value of type `Env`, listing 5, line 7). Finally, expressions evaluate to numbers (`eval` function in listing 5, line 11).

Based on this existing language definition, we now explain how to get a REPL (listing 6). `calcREPL` returns a value of type `REPL` (listing 1). As explained before, the REPL requires the definition of two functions, *handler* and *completor*. First, Calc's handler is shown in listing 6 (lines 7-15). It takes the user's input as a parameter and tries to parse it. If the parsing phase is successful, it proceeds to execute the parsed command and returns a `text` (listing 1, line 8) result (line 11, listing 6). Otherwise, if there is a parsing error, the function (`calcHandler`) returns an error message with an empty result (line 14, listing 6). Second, `calcCompletor` implements a straightforward completion function. It iterates over all the variables stored in the current environment (`env`, line 5 listing 6), and returns the set of variables that match with the `prefix` parameter. Finally, line 22 (listing 6), we construct and return a value of type `REPL` containing the functions mentioned above (`calcHandler` and `calcCompletor`).

Note that code in listings 5 and 6 is entirely independent of Bacatá. Therefore, all the code can be reused outside the notebook environment. For instance, the concrete syntax definition, the evaluation, the execution, and the `REPL` functions may all be used in a standalone IDE for Calc.

■ **Listing 5**  Calc's grammar definition using Rascal's built-in formalism.

```
1  module Syntax                          1  syntax Exp = Id
2  extend lang::std::Id;                  2            | Num
3  extend lang::std::Layout;              3            | left Exp "*" Exp
4                                         4            > left Exp "+" Exp;
5  syntax Cmd                             5
6      = Id "=" Exp                       6  alias Env = map[str, int];
7      | Exp;                             7
8                                         8  tuple[int, Env] exec(Cmd cmd, Env env) { ... }
9  lexical Num                            9
10     = [\-]?[0-9]+;                     10 int eval(Exp exp, Env env) { ... }
```

In listing 7, we detail how to generate a Jupyter language kernel with Bacatá. We use the `REPL` function `calcRepl` defined in listing 6. We first create a `Kernel` value, consisting of the language's name, the project's path, and the `REPL`'s function qualified name. `Bacata`'s function (line 4 in listing 7) returns a `Notebook` value, and as a side effect, it generates a *kernel.json* file (shown in appendix E and listing 11). The `Notebook` value may be used to start Jupyter's notebook server within the same session. Alternatively, it can also be started from the command-line outside Rascal Eclipse.

### 5.1  Input Cells: Syntax Highlighting

Jupyter's cell editor is based on the CodeMirror editor [7], thereby for syntax coloring, Jupyter uses CodeMirror *modes*. They are like so-called "Textmate grammars" [38] that are used by editors such as Textmate, VS Code, SublimeText, and many others. Listing 8 shows the definition of the `Mode` type. A `Mode` is defined by a name and a list of states, and each state has a name and several rules that apply to each state. Finally, a `Rule` defines a regular expression that matches a substring and assigns a list of tokens that determine its visual appearance. After a rule has matched, it may transit to another state via the `next` property. There are two optional Booleans `indent` and `dedent`. They are responsible for controlling auto-indentation in block constructs.

As described in section 2.4, syntax-highlighting is one of the language-specific features. Bacatá allows language engineers to describe and generate CodeMirror modes in an automatic or manual fashion. The first approach is a built-in feature of Bacatá. It takes the language's grammar and analyzes it to generate simple modes for

keyword highlighting using reflection. The second approach is by manually defining a `Mode` and sending it as a parameter to the `bacata` function. A simple mode for the Calc language is implemented as follows:

```
Mode calcMode =
  mode("Calc", [state("ini ", [rule("[0-9]+", ["number"]), rule("[a-zA-Z][a-zA-Z0-9_]*", ["variable"]) ]) ]) ;
```

■ **Listing 6**  Calc's REPL.

```
1  module Repl
2  import Syntax;
3
4  REPL calcREPL() {
5    Env env = ();
6
7    Result calcHandler(str line) {
8      try {
9        Cmd cmd = parse(#Cmd, line);
10       <n, env> = exec(cmd, env);
11       return text("<n>", []);
12     }
13     catch ParseError(loc l):
14       return text("", [message("Parse error", l)]);
15   }
16
17   Completion calcCompletor(str prefix)
18     = <pos, [ x | x ← env, startsWith(p, x) ]>
19     when /<p:[a-zA-Z]*$/ := prefix,
20       pos := size(prefix) - size(p);
21
22   return repl(calcHandler, calcCompletor);
23 }
```

■ **Listing 7**  Rascal's interactive session.

```
1  > k = kernel("Calc", |project://Calc|,
2      "Repl::calcRepl");
3  >> ...
4  > nb = bacata(k);
5  >> ...
6  > nb.serve();
7  The notebook is running at:
8    |http://localhost:8888|
```

■ **Listing 8**  Syntax Mode data type.

```
1  data Mode
2    = mode(str name, list[State] states);
3
4  data State
5    = state(str name, list[Rule] rules);
6
7
8
9  data Rule
10   = rule(str regex, list[str] tokens, str next = "",
↪      bool indent = false, bool dedent =
↪      false);
```

This mode defines a state with two rules, one for numbers and the other one for variables. To create a Calc notebook using the `calcMode` mode, one can call the `bacata` function as shown below:

```
bacata(k, calcMode);
```

## 5.2  Output Cells: Interactive Visualizations

Likewise, rich media output is also a language-dependent feature offered by notebook platforms (section 2.4). Jupyter's notebook interface runs in the browser, so this allows cells to contain arbitrary HTML/CSS/JS widgets, beyond the plain output shown in listing 6 (line 11) for the Calc language. Bacatá supports fully interactive, stateful GUIs as output cells. Said support is achieved through the integration between Bacatá and Salix [65], Rascal's web UI framework. Salix offers support for standard HTML and SVG elements, and integration with several graph frameworks and chart frameworks such as DagreJS [12] and Google Chart [15].

Salix emulates *Elm*'s architecture [11]; it is divided into three parts, namely *model*, *update*, and *view*. In Salix, these three pieces are encapsulated as an `App[&T]` type, where `&T` represents the application data *model*. An `App` encloses a *view* to draw UIs, and the *update* function to update the application's model when the user triggers an event. The way Bacatá uses Salix Apps is by allowing its usage as a REPL output. To achieve this, we extend the definition of the Result type (listing 1) as follows:

```
data Result = ...
  | app(App[&T] app, list[Message] messages);
```

As a result of extending the Result type definition, a REPL can return fully functional stateful output cells, leveraging all UI features of Salix.

Now we are going to extend our Calc language with an example expression debugger. This debugger is used as a way of interactively debugging variables values and see the effect of changing them in the evaluation of expressions. The following code snippet reflects the required changes to integrate the expression debugger in the current implementation of the Calc language. First, we add a new production rule to the Cmd non-terminal that triggers the debugging visualization as follows:

```
syntax Cmd = ...
  | "show" Exp;
```

As a result of the previous change, when the user types and executes the expression **show** x + y, the resulting output cell will contain a debugger of the expression x + y.

Listing 12 in appendix E shows the implementation of the expression debugger for the Calc language using Salix. The application model for this debugger is the environment Env and the Msg type, binds the unique event, which keeps track of changes in the variable's value. In listing 12 (Lines 3 to 20) we show a function named expDebugger that takes as arguments an expression Exp and an environment Env and produces a Salix application (App[&T]). This type encapsulates three functions, namely, init, view, and update. The first one initializes the application model. The second function takes the current environment and draws the UI based on that information. The UI shows a textual representation of the expression, including its computed value. Also, for each variable in the environment, the UI creates a label and a slider. The last function is responsible for updating the model.

Finally, to complete our expression debugger, we must include it as part of the Calc's REPL. As said before, we want to display the debugger whenever a user executes a show command. To achieve this, we added the following **if**-statement just after parsing the user's input code (listing 6, line 9).

```
Cmd cmd = ...
if (( Cmd)`show <Exp e>` := cmd) {
  return app(expApp(e, env), []);
}
```

The last **if**-statement uses Rascal's concrete syntax pattern matching to check whether cmd is a show command or not. If cmd is a show command, it binds e to the argument expression. If the match succeeds, it returns a value of type Result (using the app constructor) that encapsulates the Salix application (listing 12 and appendix E).

Figure 6 displays a Calc notebook, including a debugging interface (output cell 3). In this notebook, a user has defined two variables, x and y. Then a show command was executed to debug the effect of changing the value of the current variable bindings on the expression `2 * y`. When the user changes the slider for `y`, the new result is simultaneously updated.

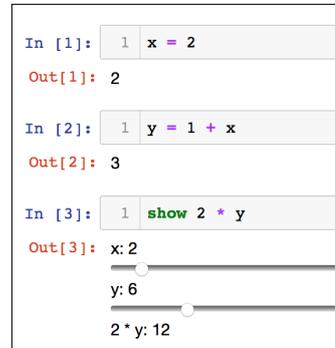

**■ Figure 6** CALC interactive debugger.

## 6    Case Studies

We have used Bacatá to generate notebook interfaces for different languages, namely, Halide*, SweeterJS, and QL. These languages were implemented using Rascal and are available on GitHub [71, 73].

### 6.1 Halide*

Halide [51] is an embedded language for image processing and computational photography. Bacatá requires a grammar and a REPL written in Rascal to create a notebook. However, we did not have a Halide grammar nor a REPL in Rascal. Therefore, we implement Halide*, a Halide grammar, and a REPL that makes the language notebook-friendly. Halide* captures a subset of the Halide language. It is important to highlight that the notebook way of working influences the design of the Halide*'s grammar and the REPL. Also, the current implementation was written in Rascal, but it could have been done in Java or using other LWB. Halide* was designed for Océ, a Canon company that develops, and manufactures printing and copying hardware. As part of their development process, one of their needs is the construction of digital image processing algorithms. In this process, there are people with different backgrounds (e.g., mathematicians, physicists, electrical engineers) not necessarily with a background in computer science. However, most of them were already familiar with a notebook way of working. They wanted a mechanism to implement said algorithms in a notebook environment that speeds up their development cycle.

Halide* divides the program into three different categories, namely, data loading, algorithms, and execution. The data loading category includes the language constructs used to load data into buffers (e.g., images, arrays); the algorithmic category describes the data transformation the user wants to express algorithmically (e.g., blur, gradient); finally, the execution category takes the data and apply algorithms over it.

Halide* generates, compiles, and executes native Halide source code; the Halide compiler is responsible for the compilation and execution steps. We introduced some syntactic sugar to the Halide grammar to detect language constructs using the categories mentioned above. Mainly, we added function wrappers to differentiate between main functions, image pipeline definitions, compilation strategies (e.g., ahead of time or just in time compilation), and execution. Halide*'s cell execution is performed through the REPL in two steps. First, we compile Halide* code into Halide code, then Bacatá delegates the compilation and execution process to $g++$. Bacatá intercepts those results, parses them into HTML, and then displays them as output cells.

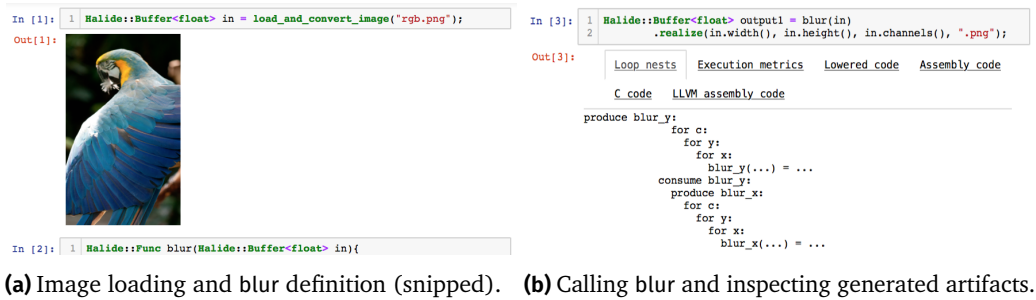

**(a)** Image loading and blur definition (snipped).  **(b)** Calling blur and inspecting generated artifacts.

■ **Figure 7**  Halide notebook.

A prototypical session using the Halide* notebook is shown in figure 7. It highlights the visualization of multimedia results and inspection of artifacts generated by the compiler. In figure 7a, the user loads a png image (as shown in the output cell [1]). Then a `blur` function is defined in the input cell [2], which does not produce an output but is now available for use. Then, in figure 7b, the `blur` function is invoked on the input image `in`. The result shows a tabbed interface built to inspect loop nesting, execution metrics, lowered code, assembly code, C code, and LLVM assembly code. Alternatively, the resulting image can be shown.

### 6.2 SweeterJS

SweeterJS [10] is a framework for language extensions of JavaScript (ECMAScript 5), and it is used to teach source-to-source transformations (desugaring) using Rascal. SweeterJS was already implemented as a Rascal language, although it did not have a REPL. Therefore, we created a SweeterJS's REPL that mostly reuses the existing SweeterJS IDE. To illustrate the benefits of notebooks

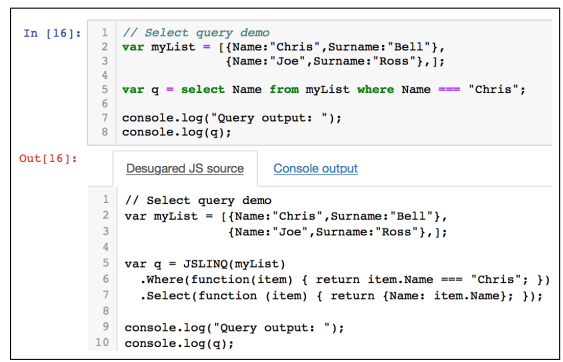

■ **Figure 8**  SweeterJS notebook.

from the language engineering perspective, we have generated a notebook interface for SweeterJS. The notebook interface allows students to experiment with the language by writing, executing, and transforming SweeterJS's code snippets. Therefore, students obtain the computed result and the desugared version (ECMAScript 5) code.

Figure 8 shows a SweeterJS notebook that contains as input some JavaScript code with an SQL-like query expression (In[16], line 5). Furthermore, it also shows the result of executing the desugared version of the code and the desugared code itself. For instance, the query expression (line 5 in the input cell) is transformed into a JSLINQ query constructor (Out[16], lines 5-7).

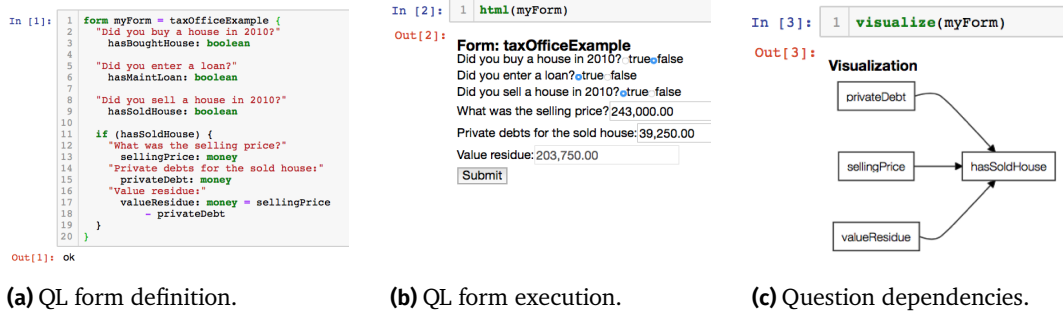

**(a)** QL form definition.  **(b)** QL form execution.  **(c)** Question dependencies.

■ **Figure 9**  Tax filing questionnaire using a QL notebook.

### 6.3 Questionnaire Language (QL)

QL is a DSL for defining interactive questionnaires that has been used to benchmark and evaluate LWBs [13]. In fact, QL is interesting from the notebook metaphor perspective since QL specifications define interactive GUI forms. There is already a Rascal implementation of QL [64], so like we did for SweeterJS, we reused the existing language definition and IDE for QL to build on top of that a REPL. A questionnaire consists of a *form* that may contain one or more *questions*, and each question has a *type*. There are three different types of questions, namely labeled, conditional, and computed. QL programs in a notebook environment can be visually represented as interactive HTML forms implemented using Salix. Besides, the QL notebook supports the visualization of control dependencies between questions, which is a valuable feature for questionnaire designers to understand the conditional logic of a questionnaire.

Figure 9 shows a QL notebook for a simple tax filing questionnaire. The user first defines a questionnaire `myForm` using the `form`-command (figure 9a). Then, in figure 9b, the form is rendered as an interactive HTML widget using the `html` command. Note that the output is a fully working questionnaire, as if we have deployed it in a production environment. Thus, this supports testing interactive questionnaires at design time. Alternatively, to understand the conditional logic of a form, the user can visualize the control dependencies using the `visualize` command (figure 9c).

### 6.4 Effort

To assess Bacatá's flexibility for creating Jupyter language kernels, we compare the number of Source Lines of Code (SLOCs) that are Bacatá independent versus the number of SLOCs required to define the notebook. These results are shown in table 1.

The Calc language (section 5) is included as a baseline. On the one hand, the reused code that is independent of Bacatá is the grammar definition and the `exec` and `eval` functions. On the other hand, the notebook specific code includes the definition of the REPL and the expression debugger.

The Halide* implementation differs from the other case studies because we reverse-engineered the Halide language to design Halide* and make it notebook-friendly. The

■ **Table 1**  Number of reused SLOCS and notebook-specific SLOCS.

| Language | Reused SLOC | Notebook SLOC |
|---------|-------------|---------------|
| Calc | 37 | 50 |
| Halide* | 51 | 647 |
| SweeterJS | 579 | 162 |
| QL | 771 | 120 |

amount of notebook specific code is higher than the reused code because it required a reengineering process of an existing non-notebook-friendly language. It required changes in several language components. The syntax definition is the only completely reusable component. When implementing the Halide* notebook, we have found some language-specific design considerations that should be addressed to change an existing language into a notebook-friendly language (section 7).

In the remaining case studies, namely SweeterJS and QL, the ratio between reused and notebook specific code is much higher. In the case of SweeterJS, the reusable code includes JavaScript's syntax definition, language extensions for state machines, queries, and a variant of HAML [66], and the required transformations to desugar language extensions to vanilla JavaScript.

In addition, the REPL for QL was already defined and it included a Salix visualization, so that we could reuse it. The same holds for other language components such as syntax definition, name resolution, and type checking. The new code includes the code for the REPL and the control-dependency visualization. However, this code can also be used outside the notebook environment.

Table 1 shows that creating a language kernel for Jupyter notebook using Bacatá requires limited effort. Its main requirement is a REPL definition, which in some cases is as simple as wiring some existing components together as we did with QL and SweeterJS. In contrast, other languages may require more profound language modifications due to the different execution models. However, even in those cases, language engineers benefit from the Jupyter's protocol implementation.

Bacatá can be used across several domains by various DSLs. In this paper, we used it for four languages, namely Calc, Halide*, SweeterJS, and QL. Therefore, we can conclude that Bacatá is functional and effectively applicable to various domains and languages. However, as discussed above, the benefit of using Bacatá may differ between domains and languages. For instance, the Halide* notebook required more effort due to the language reverse-engineering and REPL development. While, in Calc, SweeterJS, and QL, we reused the existing DSL machinery.

## 7  Discussion: Limitations and Open Problems

**DSL Design Limitations**  As mentioned in section 6.4, we have identified some DSL design considerations to make a DSL notebook-friendly. First, DSLs should support several execution entry points to avoid writing full programs in a single cell. For instance, in the Halide* notebook, we had to split up the language to perform separate operations (e.g., data loading and visualization, algorithm specification and execution) in different cells. Hence to support new entry points, we defined new language constructs via language extensions. Second, after adding support to several language entry points, we need to define a mechanism to enable the definition of programs among several cells using those entry points. For instance, the Halide* notebook keeps track of intermediate step results (e.g., data loading, algorithm definition) to use them when the user requests the execution of an algorithm.

**Notebooks and Spreadsheets**  Spreadsheets are considered the most successful end-user programming environment [20], and they have millions of users [59]. A spreadsheet is a tabular form that contains one or multiple sheets, and each sheet containing a grid of cells. Both spreadsheets and notebooks target the same audience, end-user programmers. To highlight their main differences, we defined eight features. The result of the comparison is presented in table 6.

**Layout.** Both spreadsheets and notebooks are cell-based. However, spreadsheets use a grid layout system with rows and columns (two dimensional), while notebooks have a system based solely in rows (one dimensional).

**Language.** Notebooks are language-independent, while spreadsheets are monolingual. In other words, different languages can be used through a notebook interface, whilst spreadsheets have a fixed language.

**Cell input.** As shown in figure 3, input cells in a notebook are either prose or code. A code cell in a notebook contains any statement or expression defined in the language, while spreadsheets are formula-based.

**Execution.** Both technologies have different technology modes. On the one hand, notebooks execute cells based on a linear and sequential model. On the other hand, spreadsheets use a non-sequential data flow model.

**Evaluation.** There are two types of evaluation, namely, upon request and immediate. Notebooks support cell evaluation upon request (manual execution). In contrast, spreadsheets support two evaluation mechanisms, *upon request* and *immediate*. The immediate evaluation updates cells' values when any of the values involved in the formula changes based on the dependency network.

**Persistence.** Both notebooks and spreadsheets are persisted as documents in the file system. However, there is a big difference. A notebook document can only contain one notebook, while a spreadsheet document can hold several sheets.

**Cell content.** The cell's content is the values users can see when they open either a notebook or a spreadsheet. On the one hand, notebook documents display all the information in the notebook, namely, documentation, input, and output cells. Instead, spreadsheets solely show the cell's output value.

**Input code display.** The input code in notebooks is visible for the end-user. In contrast, the input code is hidden in spreadsheets because they only display the output of the cell's content (*Cell content* feature).

**Cell references.** Notebooks support symbolic references, which means that users can reuse previous cells based on name bindings. On the other hand, spreadsheets support positional references. In other words, within a spreadsheet, end-users can reuse cell outputs based on the cell's position.

Further research is needed to explore integrating both the notebook and spreadsheet paradigms. In particular this would involve investigating how the linear narrative structure of notebooks can be reconciled with the data flow orientation of spreadsheets, and how notebooks can be made more "live" as in live programming environments.

**Summary.** In this section, we identified DSL design limitations for creating notebook-friendly languages, namely, support for several entry points of evaluation and program definition through several cells. Notebooks and spreadsheets are tailored to end-users, and both are cell-based technologies. However, there are several differences between them (table 6). We consider that there might be some languages that benefit more from a notebook interface than from a spreadsheet interface, and vice versa. In sum, not all languages fit the spreadsheet or notebook paradigm, but combining both approaches could prove very valuable.

## 8  Related Work

In the context of software language engineering, tooling support is vital to increase and improve the adoption of language-oriented programming. Bacatá contributes to the research line on program environment generation [4, 6, 13, 19, 31, 53, 62].

Fowler [14] popularized the term LWB. In one of his essays, he described the foundations of language-oriented programming, its benefits, and drawbacks. He also explains the critical role of IDE tool support in language-oriented programming. Primarily, tool support for DSLs may reduce the learning curve and boost their viability.

LWBs make use of meta-languages and meta-programming techniques that reduces the costs of building DSLs and appropriate tooling. Our work aims to follow the same trend by offering a mechanism of generating a computational notebook interface based on a language specification. Mainly, notebooks offer a different GUI for interacting with code and documentation. This interface does not clash with the more traditional GUIs such as IDEs or text editors, but it offers an enhanced interface mainly for end-user programming [35] and exploratory programming [2, 29, 56].

Interactive computing has emerged as another software development paradigm. So far, researchers have highlighted its importance and benefits for programming related tasks [9, 42]. Cook [9] highlights the main benefits of using interactive programming in software development. He illustrates the main differences between interactive and non-interactive programming. Likewise, Nagar [42] presents a case study about using Python in an interactive computing setting. Mainly, he highlights the importance and value of being able to experiment with code; mostly, the capability of executing commands and expressions, and its output. Also, the impact it has on the programming language learning curve for end-users.

Computational notebooks integrate different research areas such as literate programming, interactive computing, and end-user programming (e.g., spreadsheets). However, their main foundation is literate programming [34, 52, 57]. The basis of a computational notebook is the capacity of writing executable source code and narrative text. Also, the ability to document and explain results using multimedia formats such as charts and visualizations.

Moreover, one of the benefits of using notebooks is its sharing capabilities [60]. Turner et al. [69] explored them as a mechanism for supporting cooperative work and sharing information with non-technical staff. With Bacatá we aim at bringing DSLs closer to end-user programmers through a notebook interface.

Currently, there are several studies tackling usability, cognitive, and reproducibility aspects of computational notebooks. For instance, in reproducibility [36, 47, 48, 58], education [44], exploratory programming [2, 18, 29, 30], documentation [76], and data journalism [78]. However, no attention has been paid from the language engineering point of view. There is some work about interactive DSL usage, e.g., domain-specific debuggers [3] and live DSLs [54, 68, 70].

## 9  Conclusions & Future Work

Computational notebooks offer a different GUI for interacting with prose, executable source code, and interactive feedback. Contrary to traditional IDE and text editors, notebooks focus on a different way of working focused on computational storytelling and end-user programming (e.g., exploratory programming and data science). To better understand computational notebooks and their features, we conducted a FODA in which we studied 16 notebook platforms. Afterwards, we created a feature model where we identified language-dependent and language-independent features.

In this paper we implemented our approach in the Jupyter context. In general, developing a new language kernel for Jupyter requires much effort. In the context of DSL's, it is even more expensive because their design and implementation cycle is different from GPLs. Thus, we introduce Bacatá, a language-parametric kernel generator for Jupyter notebooks, which reuses existing language components. Then, implementing a notebook interface for a new language becomes a matter of writing a few lines of code that wire language components together as a REPL.

Moreover, we present Bacatá's architecture and how we implement it within the Rascal LWB. In addition to the default features offered by Jupyter, we integrate Rascal's web-based GUI framework (Salix) to support fully interactive output cells.

We use Bacatá-Rascal to define language kernels for four languages, namely Calc, Halide*, SweeterJS, and QL. We used the last three as case studies. As a result, we find that embedding a language into a notebook setting may have an impact on its design. In section 6.4, we compare the number of reused SLOCs versus the number of notebook-specific SLOCs. All the needed effort is focused on language-specific tasks and not in notebook-specific tasks.

In section 7, we compared spreadsheets and notebooks. Both technologies are cell-based, yet there are several differences. Thus, we consider that not all languages fit the spreadsheet or the notebook metaphors. More research is needed in this direction, to be able to classify which languages fit each metaphor. As an on-going part of this project, on the one hand, we plan to study in more detail what are both the limitations and consequences of embedding a DSL into a notebook ecosystem. Moreover, we want to evaluate if it is feasible to embed visual languages into this setting. We also plan to consolidate Bacatá's interface (ILanguageProtocol) with Microsoft's Language Server Protocol [37]. This consolidation would allow language engineers to implement a single interface once and for all. Finally, it would be interesting to explore if more features from the FODA (section 2.4) can be automatically generated to work with notebooks such as formatting, folding, polyglot kernels, and different execution modes.

## A  Computational Notebook Tools

■ **Table 2**  List of tools studied for the Feature-oriented domain analysis on computational notebook platforms.

| Tool name | URL | Access Date |
|---|---|---|
| Jupyter | https://jupyter.org/ [33] | 2019-09-22 |
| Knitr | https://yihui.name/knitr/ [39] | 2020-01-04 |
| Burrito | Research paper [16] | - |
| GraphPad Prism | https://www.graphpad.com/scientific-software/prism/ | 2020-01-04 |
| Apache Zeppelin | https://zeppelin.apache.org | 2019-09-22 |
| Observable | https://observablehq.com/ | 2019-09-22 |
| Iodide | https://alpha.iodide.io/ | 2019-09-22 |
| Distill | https://distill.pub/ | 2019-09-22 |
| Codestrates | Research paper [50] | - |
| Maple | https://maplesoft.com/products/Maple/ | 2020-01-04 |
| Azure Databricks | https://azure.microsoft.com/en-us/services/databricks/ | 2019-09-22 |
| Google Collaboratory | https://colab.research.google.com/ | 2019-09-22 |
| R Markdown | https://rmarkdown.rstudio.com/ [77] | 2019-09-22 |
| MATLAB | https://nl.mathworks.com/products/matlab/live-editor.html | 2019-09-22 |
| Mathematica | https://www.wolfram.com/mathematica | 2019-09-22 |
| Sage | http://www.sagemath.org/ [63] | 2019-09-22 |

**Table 3** Computational notebooks features (●, full support; ◐, limited support).

| | | | Apache Zeppelin | Azure Databricks | Burrito | Codestrates | Distill | Graphpad Prism | Google Colaboratory | Iodide | Jupyter | Knitr | Maple | Mathematica | MATLAB | Observablehq | R markdown | Sage |
|---|---|---|---|---|---|---|---|---|---|---|---|---|---|---|---|---|---|---|
| Editor | Editing | Free-form | ● | ● | | ● | | | ● | ● | ● | ● | | ● | ● | ● | ● | ● |
| | | Spreadsheet | | | | | | ● | | | | | | | | | | ● |
| | Syntactic services | Highlighting | ● | ● | | ● | | | ● | ● | ● | ● | | ● | ● | ● | ● | ● |
| | | Completion | ● | ● | | ● | | | ● | ● | ● | ● | | ● | ● | ● | ● | |
| | | Formatting | ● | ● | | ● | | | | ◐ | ● | | | | | | ● | |
| | | Folding | ● | ● | | ● | | | | | ● | | | | | | ● | |
| | Keyboard shortcuts | | ● | ● | | ● | | | ● | ● | ● | | | ● | ● | ● | ● | ● |
| | Line numbers | | ● | ● | | ● | | | ● | ● | ● | | | ● | | | ● | |
| | Comments | | ● | ● | | ● | ● | | ● | ● | ● | ● | | ● | ● | ● | ● | ● |
| | Prose | | ● | | | ◐ | ● | ● | | ◐ | ● | ● | | ● | | | ● | |
| | Code | Monolingual | ● | ● | | ● | | | ● | ● | ● | ● | ● | ● | ● | ● | ● | ● |
| | | Polyglot | ● | ● | | ● | | | | ● | ● | | | | | | ● | |
| | | Interactive | ● | ● | ◐ | ◐ | | | | ● | | | ● | ● | | ● | ● | |
| | Execution | Batch | ● | ● | | ● | | | | | ● | | ● | | | | ● | |
| | | Background | | ● | ● | ● | | | | ● | | | ● | ● | | | ● | |
| | | Async. tasks | | ● | ● | ● | | | | ● | | | | | | | ● | |
| | | HTML widgets | | ● | | ● | ● | | | ● | ● | ● | | ● | | | ● | |
| | Rich output | Plots | ● | ● | ● | ● | ● | ● | ● | ● | ● | ● | ● | ● | ● | ● | ● | ● |
| | | Multimedia | ● | ● | ● | ● | ● | | ● | ● | ● | ● | ● | ● | ● | ● | ● | ● |
| | | Tables | ● | ● | | ● | ● | | | ● | ● | ● | ● | ● | ● | ● | ● | ● |
| | Plain output | | ● | ● | | ● | | | | ● | ● | | | ● | | | ● | ● |
| VCS | Cell | | ● | | | ● | | | | ◐ | | | | | ● | | | |
| | Document | | ● | ● | | ● | | | | | | | | | | | ● | |
| Text editor | | | | | | ● | | | | | | | | | | | ● | |
| Licensing | Academic | | | | | | ● | | | | | | | | | | | |
| | Open source | | ● | | ● | ● | ● | | ● | ● | ● | ● | | ● | ● | | ● | ● |
| | Commercial | | | ● | | | | ● | | | | | ● | | | ● | | |
| Platform | Deployment | SaaS | ● | ● | | ● | ● | ● | ● | | ● | | | ● | ● | ● | ● | ● |
| | | Standalone | ● | | | ● | | ● | | ● | ● | ● | ● | ● | ● | | ● | ● |
| | Extensibility | 3rd party integrations | ● | ● | | ● | | | | ● | ● | | | | | | ● | |
| | | Extensions | ● | ● | | ● | ◐ | ◐ | | ● | ● | ◐ | | ● | | | ● | ● |
| | Programming lang. support | Single | ● | | | ● | | | | | | ● | ● | ● | ● | ● | ● | ● |
| | | Multiple | ● | ● | | ● | | | | ● | ● | | | | | | ● | |
| | Shareability | Offline | ● | ● | | | | | | ● | ● | ● | ● | ● | ● | | ● | |
| | | Online | ● | ● | | ● | | ◐ | | ● | ● | | | ● | | ● | ● | ● |
| | Reproducibility | Document | ● | ● | | ● | ● | | | ● | ● | | | ● | | | ● | ● |
| | | Online | ● | | | | | | | | | | | | | | ● | |

### B  Wire protocol

In this Appendix, we present implementation details on how does Bacatá implements the Jupyter's wire protocol. Table 4 shows the different sockets involved in the protocol. Moreover, table 5 contains the list of all the messages implemented in Bacatá and the socket used to send each message.

■ **Table 4**  Socket types involved in the implementation of the wire protocol.

| Socket Type |
| --- |
| Control |
| Heartbeat |
| IOPub |
| Shell |
| stdin |

■ **Table 5**  List of messages used in Bacatá-Core to implement the wire protocol.

| Message Name | Socket Type |
| --- | --- |
| shutdown_reply<br>shutdown_request | Control |
| heartbeat_message | Heartbeat |
| display_data<br>execute_input<br>stream<br>execute_result<br>status | IOPub |
| complete_request<br>execute_request<br>history_request<br>is_complete_request<br>comm_info_request<br>inspect_request<br>complete_reply<br>execute_reply<br>history_reply<br>is_complete_reply<br>inspect_reply<br>comm_info_reply<br>kernel_info_reply | Shell |

## C  ILanguageProtocol interface

In listing 9 we present the Java definition of the `ILanguageProtocol` interface.

■ **Listing 9**   ILanguageProtocol interface.

```java
public interface ILanguageProtocol {

    void initialize(Writer stdout, Writer stderr);

    String getPrompt();

    void handleInput(String line, Map<String, InputStream> output, Map<String, String> metadata)
            throws InterruptedException;

    void handleReset(Map<String, InputStream> output, Map<String, String> metadata)
            throws InterruptedException;

    boolean supportsCompletion();

    boolean printSpaceAfterFullCompletion();

    CompletionResult completeFragment(String line, int cursor);

    void cancelRunningCommandRequested();

    void terminateRequested();

    void stackTraceRequested();

    boolean isStatementComplete(String command);

    void stop();

}
```

**D    MetaJupyterServer class**

In Bacatá-Core, the `MetaJupyterServer` class abstracts the communication layer between Jupyter and a language, while the `ILanguageProtocol` (appendix C) abstracts the language from the tools in a generic way.

■ **Listing 10**   Abstract methods of the MetaJupyterServer class.

```
1  public abstract class MetaJupyterServer {
2
3      ...
4
5      /**
6       * This method processes the execute_request message and replies with an execute_reply message.
7       */
8      public abstract void processExecuteRequest(ContentExecuteRequest contentExecuteReq, Message msg);
9
10     /**
11      * This method processes the complete_request message and replies with a complete_reply
12      * message.
13      */
14     public abstract Content processCompleteRequest(ContentCompleteRequest contentCompleteRequest);
15
16     /**
17      * This method processes the history_request message and replies with
18      * a history_reply message.
19      */
20     public abstract void processHistoryRequest(Message msg);
21
22     /**
23      * This method processes the kernel_info_request message and replies with
24      * a kernel_info_reply message.
25      */
26     public abstract Content processKernelInfoRequest(Message msg);
27
28     /**
29      * This method processes the shutdown_request message and replies with
30      * a shutdown_reply message.
31      */
32     public abstract Content processShutdownRequest(ContentShutdownRequest contentShutdownRequest);
33
34     /**
35      * This method processes the is_complete_request message and replies with
36      * a is_complete_reply message.
37      */
38     public abstract Content processIsCompleteRequest(ContentIsCompleteRequest isCompleteRequest);
39
40     /**
41      * This method creates the interpreter to be used as a REPL
42      */
43     public abstract ILanguageProtocol makeInterpreter(String source, String replQualifiedName,
44                             String... salixPath) throws IOException, URISyntaxException, Exception;
45 }
```

**processExecuteRequest (listing 10 line 8)**   This method is called when an end-user request to execute an input cell. It delegates the evaluation of the input code received as a parameter to the language's REPL.

**processCompleteRequest (listing 10 line 14)** This method is executed when an end-user request to auto-complete a fragment in the current line of code.

**processHistory (listing 10 line 20)** This method returns the list of all the previously executed commands in the current environment.

**processKernelInfoRequest (listing 10 line 26)** This method is called in the initialization of the language kernel. When an end-user creates a new notebook, Jupyter calls this method to obtain information about the language (e.g., name, version, language logo, CodeMirror mode).

**processShutdownRequest (listing 10 line 32)** This method is used to stop a language kernel.

**processIsCompleteRequest (listing 10 line 38)** This method decides whether the input cell code is complete or not. If the code is incomplete, it tells the front-end to display a continuation prompt.

**makeInterpreter (listing 10 lines 43-44)** This method returns an instance of the *ILanguageProtocol*. This instance acts as a bridge between the Jupyter's communication layer and the language. Thus, this object interacts with the language's REPL.

**E**   **Additional Calc Listings**

■ **Listing 11**   CALC's language kernel file.

```
 1  {
 2    "argv": [
 3      "java", "-jar",
 4      "/User/bacata/bacata-dsl.jar",
 5      "{connection_file}",
 6      "home:///projects/Calc",
 7      "Repl::myRepl",
 8      "Calc"
 9    ],
10    "display_name": "Calc",
11    "language": "Calc"
12  }
```

■ **Listing 12**   Expression debugger defined using Salix.

```
 1  data Msg = var(str x, str val);
 2
 3  App[Env] expApp(Exp e, Env env) {
 4    Env init ()  = env;
 5    void view(Env env) {
 6      div(()  {
 7        for (str x ← env) {
 8          text("<x>: <env[x]>");
 9          input(\type("range"), \value(env[x]), onInput(partial(var, x)));
10        }
11        text("<e>: <eval(e, env)>");
12      });
13    }
14    Env update(var(x, v), Env env) = env + (x: toInt(v));
15    return makeApp(init, view, update);
16  }
```

**F** Bacatá's Architecture – Class Diagram

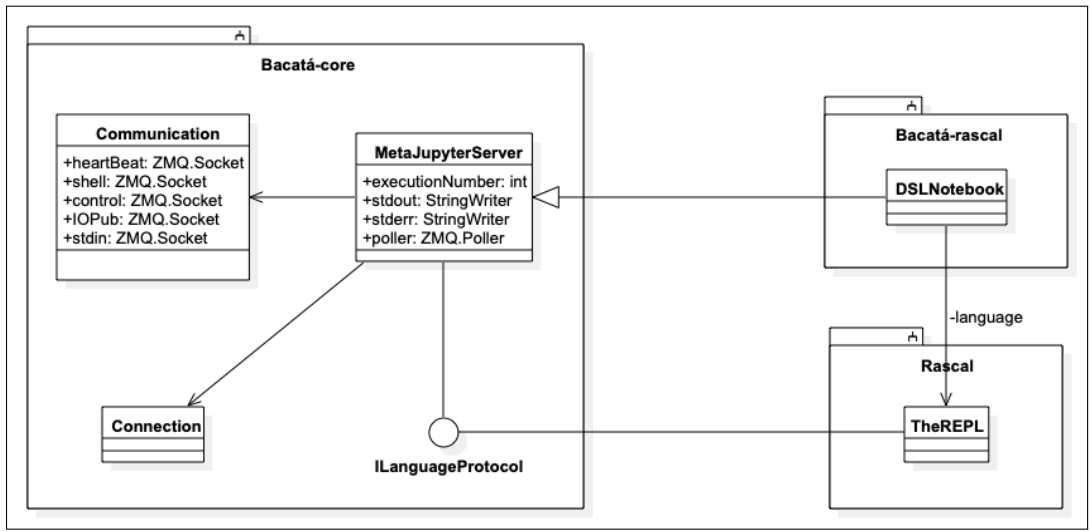

■ **Figure 10** Bacatá – Class Diagram.

**G** Comparing Computational Notebook and Spreadsheet metaphors.

■ **Table 6** Comparison between notebook and spreadsheet metaphors.

| Features | Notebooks | Spreadsheets |
|---|---|---|
| *Layout* | Grid (2D) | Rows (1D) |
| *Language* | Language-independent | Fixed language |
| *Cell Input* | Statements/Expressions | Formulas |
| *Execution* | Sequential | Data flow/Dependency network |
| *Evaluation* | Upon request | Immediately |
| *Persistence* | Single | Multiple |
| *Cell content* | Doc./Code/Output | Output |
| *Input Code* | Visible | Hidden |
| *Cell References* | Symbolic ref. | Positional ref. |

## About the authors

**Mauricio Verano Merino** is a PhD candidate in the Software Engineering and Technology group at the Eindhoven University of Technology. His research focuses on programming environment generation and user-interfaces. He aims at improving the adoption and user experience of domain-specific languages. He can be reached at m.verano.merino@tue.nl.

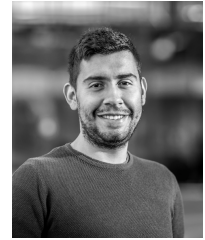

**Jurgen Vinju** is full professor of Automated Software Analysis at Eindhoven University of Technology, research group leader at Centrum Wiskunde & Informatica (CWI), and senior language engineer and co-founder of SWAT.engineering. He studies the design and evaluation of (applications of) meta programming systems to get the complexity of source code maintenance under control. Examples are metrics and analyses for quality control or debugging, and model driven engineering for code generation. For more information, see http://www.cwi.nl/~jurgenv. He can be reached at Jurgen.Vinju@cwi.nl.

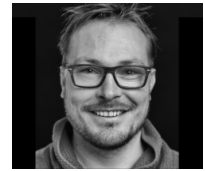

**Tijs van der Storm** is senior researcher in the Software Analysis and Transformation group at Centrum Wiskunde & Informatica (CWI), and full professor in Software Engineering at the University of Groningen. His research focuses on improving programmer experience through new and better software languages and developing the tools and techniques to engineer them in a modular and interactive fashion. For more information, see http://www.cwi.nl/~storm. He can be reached at storm@cwi.nl.

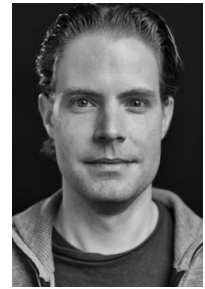